\documentclass[journal]{IEEEtran}

\IEEEoverridecommandlockouts

\usepackage[fleqn]{amsmath}
\usepackage{amssymb,graphicx,verbatim,enumitem,mdframed,mathrsfs}

\newtheorem{theorem}{\textbf{Theorem}}
\newtheorem{lemma}{\textbf{Lemma}}
\newtheorem{corollary}{\textbf{Corollary}}
\newtheorem{definition}{\textbf{Definition}}

\providecommand{\mnorm}[1]{|#1|}

\title{\LARGE \bf Statistical Privacy in Distributed Average Consensus on Bounded Real Inputs}

\author{Nirupam Gupta, Jonathan Katz and Nikhil Chopra
\thanks{This work was partially supported by the National Science Foundation through award ECCS1711554.}
\thanks{Nirupam Gupta ({\tt\small nirupam@umd.edu}) and Nikhil Chopra ({\tt\small nchopra@umd.edu}) are with the Department of Mechanical Engineering,
        University of Maryland, College Park, 20742 MD, USA}
\thanks{Jonathan Katz ({\tt\small jkatz@cs.umd.edu}) is with the Department of Computer Science,
        University of Maryland, College Park, 20742 MD, USA}
}

\begin{document}

\maketitle





\begin{abstract}                
This paper proposes a privacy protocol for distributed average consensus algorithms on bounded real-valued inputs that guarantees statistical privacy of honest agents' inputs against colluding (passive adversarial) agents, if the set of colluding agents is not a vertex cut in the underlying communication network. This implies that privacy of agents' inputs is preserved against $t$ number of arbitrary colluding agents if the connectivity of the communication network is at least $(t+1)$. A similar privacy protocol has been proposed for the case of bounded integral inputs in our previous paper~\cite{gupta2018information}. However, many applications of distributed consensus concerning distributed control or state estimation deal with real-valued inputs. Thus, in this paper we propose an extension of the privacy protocol in~\cite{gupta2018information}, for bounded real-valued agents' inputs, where bounds are known apriori to all the agents.    

\end{abstract}


\section{Introduction}
\label{sec:intro}
\emph{Distributed average consensus} algorithms (for eg. \cite{jadbabaie2003coordination, boyd2006randomized}) can be used in a peer-to-peer network by agents to reach a consensus value, equal to the average of all the agents' inputs. Some of the applications of distributed average consensus include sensor fusion \cite{olfati2005distributed}, solving economic-dispatch problem in smart grids \cite{yang2013consensus}, and peer-to-peer online voting.

Typical distributed average consensus algorithms require the agents to share their inputs (and intermediate states) with their neighbors \cite{jadbabaie2003coordination, boyd2006randomized}. This infringes the privacy of agents' inputs, which is undesirable as certain agents in the network may be passive adversarial\footnote{Passive adversarial agents follow the prescribed protocol unlike active adversarial agents, but can use their information to gather information about the inputs of other agents in the network.} and non-trustworthy~\cite{huang2012differentially,manitara2013privacy,mo2017privacy,nozari2017differentially,ruan2017secure,gupta2017privacy}.  

If the agents' inputs are integers (bounded), privacy in distributed average consensus can be achieved by relying on (information-theoretic) distributed secure multi-party computation protocols~\cite{garay2008almost} or homomorphic encryption-based average consensus~\cite{ruan2017secure, lazzeretti2014secure}. In this paper, we are interested in real-valued inputs with \emph{known} bound, as several applications of distributed average consensus such as distributed Kalman filtering~\cite{olfati2005distributed}, formation control~\cite{ren2008distributed} and distributed learning~\cite{forero2010consensus}---deal with real-valued agents' inputs.

Several proposals~\cite{huang2012differentially, mo2017privacy, nozari2017differentially}
achieve \emph{differential privacy} by having agents obscure their intermediate states (or values) by adding locally generated noise in a particular synchronous distributed average consensus protocol. Adding such local noises induces a loss in accuracy~\cite{nozari2017differentially, braca2016learning} and there is an inherent trade-off between privacy and the achievable accuracy (agents are only able to compute an \emph{approximation} to the exact average value). Schemes in~\cite{mo2017privacy, manitara2013privacy} iteratively cancel the noise added over time to preserve the accuracy of the average of all inputs. In the proposed privacy protocol, the random values added by agents to hide their inputs are correlated over space (in context of communication network) than over time, and collectively add up to zero, hence preserving the average value of the inputs. Note that differential privacy guarantees inevitably change if the agents' inputs are bounded by a value \emph{known} to all the agents. In this paper, we are interested in statistical privacy guarantee specifically for the case when inputs have a \emph{known} bound.

 Scheme in~\cite{pequito2014design} proposes re-designing of network link weights to limit the \emph{observability} of agents' inputs but every agent's input gets known to its neighbors. The scheme of Gupta et al.~\cite{gupta2016confidentiality} assumes a centralized (thus, not distributed), trusted authority that distributes information to all agents each time they wish to run the consensus algorithm.

We note that some of the above solutions~\cite{huang2012differentially, manitara2013privacy, mo2017privacy, nozari2017differentially} require \emph{synchronous} execution of the agents, whereas our privacy protocol is asynchronous (refer Section~\ref{sec:pm}). Moreover, this is the first paper, to the best of authors' knowledge, to propose a privacy protocol for distributed average consensus on bounded real-value inputs where bounds are apriori known. It is important to note that prior knowledge of inputs' bounds makes the privacy problem more challenging and renders the existing claims on differential privacy invalid.
\subsection{Summary of Contribution}
We develop on our previous works~\cite{gupta2018information,gupta2017privacy} to propose a privacy protocol that guarantees statistical privacy of honest (non-adversarial) agents' inputs against colluding passive adversarial agents in any distributed average consensus over bounded (bounds \emph{known} to all agents) real-valued inputs. In~\cite{gupta2017privacy} we proposed a general approach for achieving privacy in distributed average consensus protocols for both real-valued and integral inputs. However, the privacy guarantee in~\cite{gupta2017privacy} is weaker and uses relative entropy (KL-divergence) instead of the more standard statistical distance for privacy analysis. It is to be noted that the privacy approach in~\cite{gupta2018information,gupta2017privacy} for integral inputs is quite similar to the one proposed by Emmanuel et al.~\cite{abbe2012privacy}. However,~\cite{abbe2012privacy} only considers a complete network topology which is relaxed in our work. Moreover, we focus on real-valued inputs and thus, the privacy scheme in~\cite{abbe2012privacy} is not readily applicable. The privacy scheme in~\cite{abbe2012privacy} has been extended for privacy in distributed optimization by~\cite{gade2016private} for real-valued agents' costs (equivalent to `inputs' in our case). However, the privacy analysis in~\cite{gade2016private} does not provide any formal quantification on privacy guaranteed, and is not applicable to the case when the inputs are bounded with bounds being known apriori to all the agents. 

Our proposed protocol constitutes of two phases:
\begin{enumerate}
    \item 
In the first phase, each agent share correlated random values with its neighbors and computes a new, ``effective input'' based on its original input and the random values.

\item In the second phase, the agents run any (non-private) distributed average consensus protocol (for eg. \cite{jadbabaie2003coordination}) to compute the sum of their effective inputs.
\end{enumerate}
By design, the first phase ensures that the average of the agents' effective inputs is equal to the average of their original inputs (under a particular mathematical operator). Therefore, the two-phase approach does not affect the accuracy of the average value of the inputs. Furthermore, the privacy holds in our approach---in a formal statistical sense and under certain conditions, as discussed below---regardless of the average consensus protocol used in the second phase.
To prove this we consider the worst-case scenario where all the effective inputs of the honest agents are revealed to the colluding semi-honest parties in the second phase.

The notion of privacy is the same as that used for the case of integral inputs in our earlier work~\cite{gupta2018information}, which had been adopted from the literature on secure multi-party computation~\cite{Goldreich04}. Informally, the guarantee is that the entire \emph{view} of the colluding agents throughout the execution of our protocol can be \emph{simulated} by those agents given (1)~their original inputs and (2)~the average of the original inputs of the honest agents (or, equivalently, the average of the original inputs of all the agents in the network). This holds regardless of the true inputs of the honest agents. As a consequence, this means that the colluding adversarial agents learn nothing about the collective inputs of the honest agents from an execution of the protocol other than the average of the honest agents' inputs, and this holds regardless of any prior knowledge the adversarial agents may have about the inputs of (some of) the honest agents, or the distribution of those inputs. 
We prove that our protocol satisfies this notion of privacy as long as the set of colluding adversarial agents is not a vertex cut in underlying the communication network.

\section{Notation and Preliminaries}
\label{sec:not}

\def\Z{{\mathbb Z}}

We let 
$\mathbb{R}$ denote the set of non-negative real numbers and $frac(x) \in [0,1)$ denote the fractional part of $x \in \mathbb{R}$. For any interval $[a,b] \in \mathbb{R}$, $[a,b]^n$ denotes the set of $n$-dimensional vectors with element taking values in $[a,b]$. We rely on the following basic properties 
\begin{align*}
    frac(x + y) &= frac \left( frac(x) + frac(y) \right) \\
    frac(-x) &= 1 - frac(x), \, x \neq 0
\end{align*}

If $x$ is an $n$-dimensional vector, then $x_i$ denotes its $i$th element and $\sum_i x_i$ simply denotes the sum of all its elements. We use $1_n$ to denote the $n$-dimensional vector all of whose elements is~$1$.

\label{sub:gt}
We consider communication networks represented by  simple, undirected graphs. That is, the communication links in a network of $n$ agents is modeled via a 
graph 
$\mathcal{G} = \{\mathcal{V}, \,\mathcal{E} \}$ where
the nodes $\mathcal{V} \triangleq \{1,\ldots,n\}$ denote the agents, and there is an edge $\{i, j\} \in \mathcal{E}$ iff there is a direct communication channel between agents $i$ and~$j$.
We let $N_i$ denote the set of neighbors of an agent $i \in \mathcal{V}$, i.e., $j \in N_i$ if and only if $\{i,\,j\} \in \mathcal{E}$. (Note that $i \not \in N_i$ since $\mathcal{G}$ is a simple graph.) 


We say two agents $i, j$ are \emph{connected} if there is a path from~$i$ to~$j$; since we consider undirected graphs, this notion is symmetric. We let $p_{i,j}$ denote an arbitrary path between $i$ and~$j$, when one exists. A graph $\mathcal{G}$ is \emph{connected} if every distinct pair of nodes is connected; note that a single-node graph is connected.

\begin{definition}{(Vertex cut)}
A set of nodes $\mathcal{V}_{cut} \subset \mathcal{V}$ is a \emph{vertex cut} of a graph $\mathcal{G}=\{\mathcal{V}, \mathcal{E}\}$ if removing the nodes in~$S$ (and the edges incident to those nodes) renders the resulting graph unconnected. Then, we say that $\mathcal{V}_{cut}$ \emph{cuts}~$\mathcal{V}\setminus \mathcal{V}_{cut}$.
\end{definition} 

A graph is \emph{$k$-connected} if the smallest vertex cut of the graph contains~$k$ nodes.

\def\G{\mathcal{G}}
\def\V{\mathcal{V}}
\def\E{\mathcal{E}}

Let $\G =\{\V, \E\}$ be a graph. The \emph{subgraph induced by $\V' \subset \V$} is the graph $\G' = \{\V', \E'\}$ where $\E' \subset \E$ is the set of edges entirely within $\V'$ (i.e., $\E' = \{\{i, j\} \in \E \mid i, j \in \V'\}$). We say 
a graph $\mathcal{G}=\{\mathcal{V}, \mathcal{E}\}$ has \emph{$c$ connected components} if its vertex set $\V$ can be partitioned into disjoint sets $\V_1, \ldots, \V_c$ such that (1)~$\G$ has no edges between $\V_i$ and $\V_j$ for $i \neq j$ and (2)~for all~$i$, the subgraph induced by $\V_i$ is connected.
Clearly, if $\mathcal{G}$ is connected then it has one connected component.


For a graph $\G=\{\V, \E\}$, we define its \emph{incidence matrix}
$\nabla \in \{-1,0,1\}^{\mnorm{\mathcal{V}}\times \mnorm{\mathcal{E}}}$ (see \cite{godsil2001algebraic})
to be the matrix with $|\V|$ rows and $|\E|$ columns in which
\[
	\nabla_{i,\,e} = \left\{\begin{array}{cl}1 & \hspace*{3pt} \text{if } e = \{i,\,j\} \text{ and } i < j\\ -1 &  \hspace*{3pt} \text{if } e = \{i,\,j\} \text{ and } i > j \\ 0 &  \hspace*{3pt} \text{otherwise.}\end{array}\right. 
\]
Note that $1_n^T\cdot \nabla = 0$. We use $\nabla_{*,e}$ to denote the column of $\nabla$ corresponding to the edge $e \in \mathcal{E}$.

We rely on the following result \cite[Theorem 8.3.1]{godsil2001algebraic}:
\begin{lemma}
\label{lem:o_m}
Let $\G$ be an $n$-node graph with incidence matrix $\nabla$. Then
$\text{rank}(\nabla) = n-c$, where $c$ is the number of connected components of~$\mathcal{G}$. 
\end{lemma}

\subsection{Problem Formulation}
\label{sec:prob_f}
We consider a network of $n$ agents where the communication network between agents is represented by an undirected, simple, connected graph~$\G=\{\V, \E\}$; that  is, agents $i$ and $j$ have a direct communication link between them iff $\{i, j\} \in \E$.
The communication channel between two nodes/agents is assumed to be both private and authentic; equivalently, in our adversarial model we do not consider an adversary who can eavesdrop on communications between honest agents, or tamper with their communication\footnote{Alternately, private and authentic communication can be ensured using standard cryptographic techniques.}.

Each agent $i$ holds a (private) input~$s_i$. By scaling appropriately\footnote{Suppose each agent holds a finite real-valued input $x_i \in [0,q), \, q \in \mathbb{R}^{+}$, then  $s_i = x_i/nq \in [0,1/n)$.}, we can assume without loss of generality that $s_i \in [0, 1/n)$, where $n$ is the number of agents in the network.
We let $s = [s_1,\ldots,\, s_n]^T \in [0,1/n)^n$. 
A \emph{distributed average consensus algorithm} is an interactive protocol allowing the agents in the network to each compute the average of the agents' inputs, i.e., after execution of the protocol each agent outputs the value $\bar s = \frac{1}{n} \cdot \sum_i s_i$. The value of $n$ is assumed known to all the agents.

\def\C{\mathcal{C}}
\def\H{\mathcal{H}}

We are interested in distributed average consensus algorithms that ensure privacy against an attacker who controls some fraction of the agents in the network. 
We let $\C \subset \V$ denote the set of passive adversarial, and let $\H = \V\setminus \C$ denote the remaining honest agents.
As stated earlier, we assume the adversarial agents are passive and thus run the prescribed protocol. Privacy requires that the entire \emph{view} of the adversarial agents---i.e., the inputs of the adversarial agents as well as their internal states and all the protocol messages they received throughout execution of the protocol---does not leak (significant) information about the original inputs of the honest agents. Note that, by definition, the set of adversarial agents learns $\bar s$ (assuming at least one agent is adversarial) from the sum of the inputs of the honest agents can be computed, and so our privacy definition requires that the adversarial agents do not learn anything more than this. 


\def\view{{\sf View}}

Before giving our formal definition of privacy, we introduce some notation. Let $s_\C$ denote a set of inputs held by the agents in $\C$, and $s_\H$ a set of inputs held by the agents in $\H$. Fixing some protocol, we let $\view_\C(s)$ be a random variable denoting the view of the agents in $\C$ in an execution of the protocol when the agents all begin holding inputs~$s$.
Then:

\begin{definition}\label{def:ip}
A distributed average consensus protocol is \emph{(perfectly) $\C$-private} if for all $s, s' \in [0,1/n)^n$ such that  $s_\C=s'_\C$ and 
$\sum_{i \in \H} s_i = \sum_{i \in \H} s'_i$,
the distributions of $\view_\C(s)$ and $\view_\C(s')$ are identical.
\end{definition}

We remark that this definition makes sense even if $|C|=n-1$, though in that case the definition is vacuous since $s_\H = \sum_{i \in \H} s_i$ and so revealing the sum of the honest agents' inputs reveals the (single) honest agent's input!

An alternate, perhaps more natural, way to define privacy 
is to require that for any distribution $S$ (known to the attacker) over the honest agents' inputs,
the distribution of the honest agents' inputs conditioned on the attacker's view is identical to the 
distribution of the honest agents' inputs conditioned on their sum. It is not hard to see that this is equivalent to the above definition.

\section{Private Distributed Average Consensus}
\label{sec:pm}

As described previously, our protocol has a two-phase structure. In the first phase, each agent~$i$ computes an ``effective input'' $\tilde s_i$ based on its original input~$s_i$ and random values it sends to its neighbors; this is done while ensuring that $frac(\sum_i \tilde s_i)$ is equal to $\sum_i s_i$ (see below). In the second phase, the agents use any (correct) distributed average consensus protocol $\Pi$ to compute $\sum_i \tilde s_i$, take its fractional part, and then divide by~$n$. This (as will be shown) gives the correct average $\frac{1}{n} \cdot \sum_i s_i$.

We prove privacy of our algorithm by making a ``worst-case'' assumption about $\Pi$, namely, that it simply reveals all the agents' inputs to all the agents. Such an algorithm is, of course, not at all private; for our purposes, however, this does not violate privacy because $\Pi$ is run on the agents' \emph{effective} inputs~$\{\tilde{s}_i\}$ rather than their true inputs~$\{s_i\}$. Therefore, the privacy result holds regardless of the distributed average consensus protocol $\Pi$. 
From now on, then, we let the \emph{view} of the adversarial agents consist of the original inputs of the adversarial agents, their internal states and all the protocol messages they receive throughout execution of the first phase of our protocol, and the vector~$\tilde{s} = [\tilde{s}_1, \ldots, \tilde{s}_n]^T$ of all agents' effective inputs at the end of the first phase. Our definition of privacy (cf.\ Definition~\ref{def:ip}) remains unchanged.

The first phase of our protocol proceeds as follows:
\begin{enumerate}
	\item Each agent $i \in \mathcal{V}$ chooses independent, uniform values $r_{ij} \in [0,1)$ for all $j \in \mathcal{N}_i$, and sends $r_{ij}$ to agent~$j$.
	\item Each agent $i \in \mathcal{V}$ computes a mask
		\begin{align}
			a_i = frac\left(\sum_{j \in N_i}(r_{ji}-r_{ij}) \right), \label{eqn:i_vn}
		\end{align}
		where $a_i \in [0,1)$.
	\item Each agent $i \in \mathcal{V}$ computes  effective input
	\begin{align}
		\tilde{s}_i = frac(s_i + a_i). \label{eqn:mask_coeff}
	\end{align}
\end{enumerate}
Note that
\[
frac \left(\sum_i \tilde s_i \right)  =  frac \left(\sum_i s_i + frac\left(\sum_i a_i\right) \right) 
\]
As $\G$ is undirected, therefore
\begin{align*}
frac\left(\sum_i a_i\right) = frac\left(\sum_i \sum_{j \in N_i}(r_{ji}-r_{ij})\right) = 0
\end{align*}
Thus, 
$ frac \left(\sum_i \tilde s_i\right) = \sum_i s_i$, since $\sum_i s_i < 1$ as $s_i \in [0,1/n), \, \forall i$. Hence, correctness of our overall algorithm (i.e., including the second phase) follows.

Note that any two neighboring agents $i$ and $j$ choose values $r_{ij}$ and $r_{ji}$, respectively, independently. Agents $i$ and $j$ then transmit these values $r_{ij}$ and $r_{ji}$, respectively to each other in an independent manner as well\footnote{Agent $i$ transmits $r_{ij}$ regardless of whether it has received $r_{ji}$ or not. Same applies for agent $j$.}. Therefore, Step 1 does not require synchronicity between any two agents. Steps 2 and 3 are performed locally, and therefore synchronicity between agents is out of question. Once an agent completes the first-phase, it floods the network with this information regardless of whether any other agent has completed the first-phase or not. As every agent has prior knowledge of the total number of agents, the agents reach an agreement on the completion of the first-phase when $\G$ is connected. \emph{Hence, the first-phase is asynchronous and this implies that the proposed protocol is asynchronous if the distributed average consensus protocol in the second-phase is asynchronous.}\\
In the second-phase, the agents can use an asynchronous distributed average consensus protocol, such as the randomized gossip algorithm~\cite{boyd2006randomized}, to compute the average value of $\{n\tilde s_i\}$, which equal to $\sum_i \tilde s_i$.



\subsection{Privacy Analysis}
\label{sec:pa}

We show here that $\C$-privacy holds if $\C$ is not a vertex cut of~$\G$ under the assumptions on agents' inputs, network topology and communication links mentioned in Section~\ref{sec:prob_f}.

\label{sub:masks}

For an edge $e=\{i,j\}$ in the graph with $i<j$,
define \[b_e = frac(r_{ji}-r_{ij}).\]
Let $b=[b_{e_1}, \ldots]$ be the collection of such values for all the edges in~$\G$.
If we let $a=[a_1, \ldots, a_n]^T$ denote the masks used by the agents, then we have
\[a = frac(\nabla \cdot b) .\]
Since the $r_{ij}$ are uniform and independent in~$[0,1)$, it is easy to see that the values $\{b_e\}_{e \in \E}$ are uniform and independent in~$[0,1)$ as well\footnote{If $x$ and $y$ are two independent random variables in $[0,1)$ with at least one of them being uniformly distributed, then $z = frac(x + y)$ is uniformly distributed in $[0,1)$.}.
Thus, $a$ is uniformly distributed over the vectors in the span of the columns of~$\nabla$, which we denote by~$L(\nabla)$, with coefficients in $[0,1)$.
The following is easy to prove using the fact that 
$\text{rank}(\nabla)=n-1$  when $\G$ is connected (cf.\ Lemma~\ref{lem:o_m}):
\begin{lemma}
\label{lem:dist_a}
If $\mathcal{G}$ is connected then $a$ is uniformly distributed over all points in $[0,1)^{n}$ subject to the constraint that $frac (\sum_i a_i) = 0$. 
\end{lemma}
(A full proof of Lemma~\ref{lem:dist_a} is given in Appendix~\ref{sub:dist}.)

Since $\tilde{s}_i = frac(s_i + a_i)$, we have
\begin{lemma}
\label{lem:cond_mask}
	If $\mathcal{G}$ is connected, 
	then the effective inputs $\tilde s$ are uniformly distributed in $[0,1)^n$ subject to the constraint that $frac \left( \sum_i \tilde s_i \right) = \sum_i s_i$.
\end{lemma}
The proof of Lemma~\ref{lem:cond_mask} is given in Appendix~\ref{sub:cond}.

The above implies privacy for the case when $\C=\emptyset$, i.e., when there are no adversarial agents. In that case, the view of any agent consists only of the effective inputs~$\tilde{s}$, and Lemma~\ref{lem:cond_mask} shows that the distribution of those values 
depends only on the sum of the agents' true inputs. 
Below, we extend this line of argument to the case of nonempty~$\C$.




\label{sub:suff}

Fix some set $\C$ of passive adversarial agents, and recall that $\H=\V\setminus \C$. Let $\E_\C$ denote the set of edges incident to~$\C$, and let $\E_{\H} = \E \setminus \E_\C$ be the edges incident only to honest agents. 
Note that now the view of adversarial agents' view contains (information that allows it to compute) $\{b_e\}_{e \in \E_\C}$ in addition to the honest agents' effective inputs~$\{\tilde s_i\}_{i \in \H}$.

The key observation enabling a proof of privacy is that the values $\{b_e\}_{e \in \E_{\H}}$ are uniform and independent in $[0,1)^{|\H|}$ \emph{even conditioned on the values of~$\{b_e\}_{e \in \E_\C}$}. Thus, owing to Lemma \ref{lem:dist_a}, as long as $\C$ is not a vertex cut of~$\G$, an argument as earlier implies that the masks $\{a_i\}_{i \in \H}$ are uniformly distributed in $[0,1)^{|\H|}$ subject to $frac(\sum_{i \in \H} a_i) = frac(-\sum_{i \in \C} a_i)$ (even conditioned on knowledge of the values~$\{b_e\}_{e \in \E_\C}$), and hence the effective inputs $\{\tilde s_i\}_{i \in \H}$ are uniformly distributed in $[0,1)^{|\H|}$ subject to 
\[frac\left(\sum_{i \in \H} \tilde s_i \right) =  frac\left(\sum_{i \in \V} s_i - \sum_{i \in \C} \tilde s_i \right) \] (again, even conditioned on knowledge of the~$\{b_e\}_{e \in \E_\C}$). Since the right-hand side of the above equation can be computed from the effective inputs of the adversarial agents, the $\{b_e\}_{e \in \E_\C}$, and the sum of the honest agents' inputs, this implies:

\begin{theorem}
\label{thm:perf}
If $\C$ is not a vertex cut of $\G$, then our proposed distributed average consensus protocol is perfectly $\C$-private.
\end{theorem}

A formal proof of this theorem
is given in Appendix~\ref{sub:perf}.

As a corollary, we have

\begin{corollary}
\label{cor:f}
If $\G$ is $(t+1)$-connected, then for any $\C$ with $|\C| \leq t$ our proposed distributed average consensus protocol is perfectly $\C$-private.
\end{corollary}

In case the passive adversarial agents do form a vertex cut, in that case the proposed privacy protocol guarantees privacy of each set of honest agents that is not cut by $\C$, in the sense as formally defined\footnote{$\C$ does not cut a set of agents if that set of agents that is connected in the residual graph after removing $\C$ and $\E_\C$.}. Alternately, for a set of honest agents $\H' \subset \H$ that is not cut by $\C$ the adversarial agents can deduce anything about their collective inputs $\{s_i\}_{i \in \H'}$ other than their sum $\sum_{i \in \H'}s_i$. (refer~\cite{gupta2018information})



\section{Illustration}
\label{sec:illus}
\begin{figure}[htb!]
\begin{center}
	\includegraphics[width=0.25\textwidth]{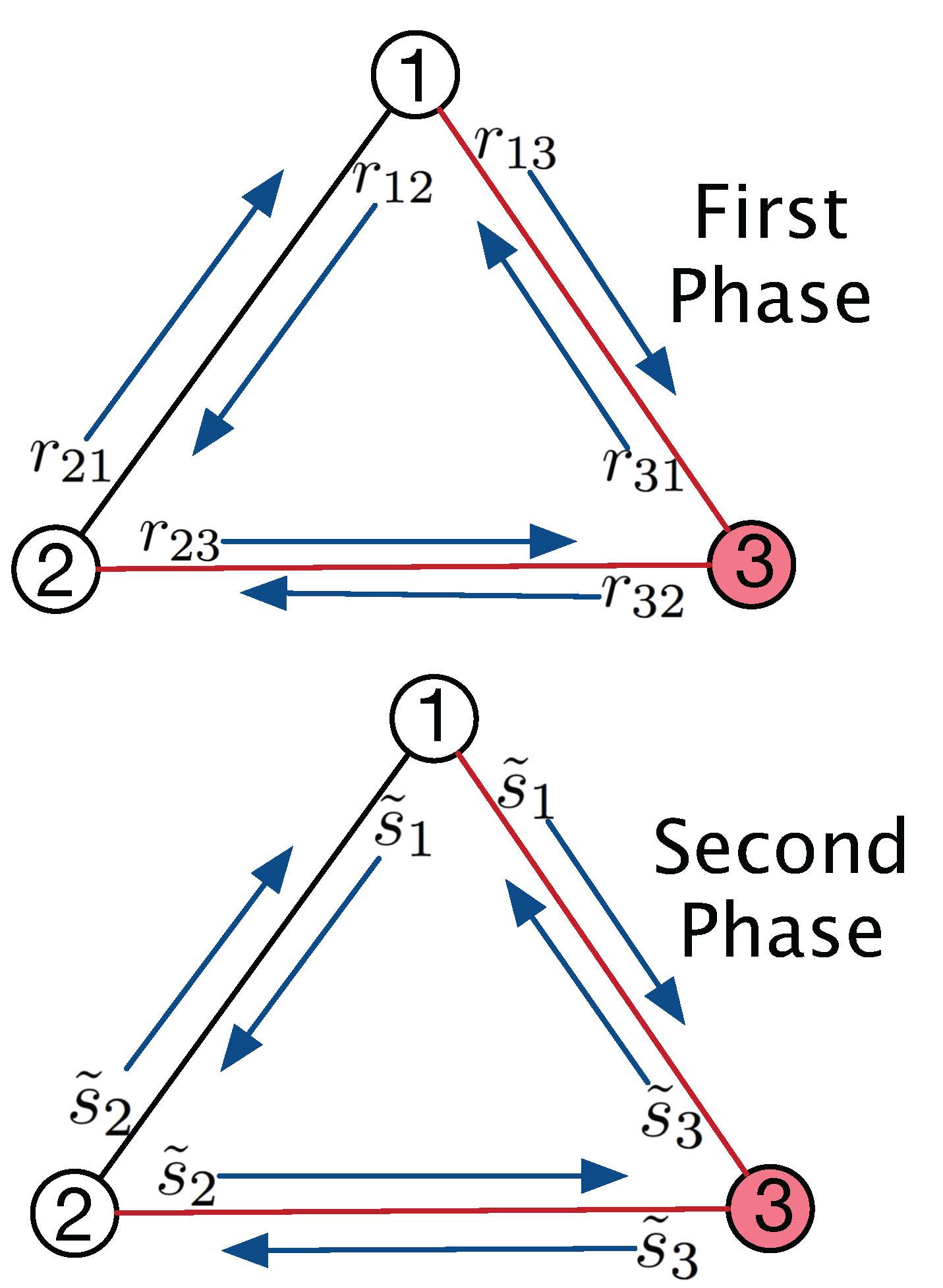}    
	\caption{\footnotesize{ Arrows (in blue) show the flow of information over an edge. }}
	\label{fig:illust}
\end{center}
\end{figure}

To demonstrate our proposed distributed average consensus protocol we consider a simple network of $3$ agents with $\mathcal{V} = \{1,\, 2, \, 3\}$ and $\mathcal{E} = \left\{ \{1,\,2\}, \, \{1,\,3\}, \, \{2,\,3\} \right\}$, as shown in Fig. \ref{fig:illust}.  
Let $s_1 = 0.1$, $s_2 = 0.2$ and $s_3 = 0.15$.

In first phase, the agents execute the following steps
\begin{enumerate}
        \item 
        As shown in Fig. \ref{fig:illust}, all pair of adjacent agents $i$ and $j$ exchange the respective values of $r_{ij}$ and $r_{ji}$ (chosen independently and uniformly in $[0,1)$) with each other. Consider a particular instance where: $r_{12} = 0.1, \, r_{21} = 0.5, \, r_{23} = 0.7, \, r_{32} = 0.4, \, r_{31} = 0.3 \text{ and }r_{13} = 0.8$. 
        \item
            The agents compute their respective masks, 
            \begin{align*}
            	a_1 &=  frac\left((r_{21} - r_{12}) + (r_{31} - r_{13}) \right) = 0.9
            \end{align*}
            Similarly, $a_2 = 0.3$ and $a_3 = 0.8$. (One can verify that $frac(a_1 + a_2 + a_3) = 0$.)
        \item 
            The agents compute their respective effective inputs,
            \begin{align*}
            	& \tilde{s}_1 = frac(s_1 + a_1)  = frac(0.1 + 0.9) = 0.0
            \end{align*}
            Similarly, $\tilde{s}_2 = 0.5$ and $\tilde{s}_3 = 0.95$.
\end{enumerate}

After the first phase, each agent uses a (non-private) distributed average consensus protocol $\Pi$ (an instance shown in Fig. \ref{fig:illust}) in the second phase to compute $(1/3)\sum_{i}\tilde{s}_i$ (it can be easily to verified that $frac(\sum_{i} \tilde{s}_i) =  \sum_{i} s_i = 0.45$).

Let $\C = \{3\}$ and so, $\E_{\C} = \{\{1,\,3\}, \, \{2,\,3\}\}$. It is easy to see that $\C$ does not cut the graph $\G$ and therefore, for any pair of inputs $s_1 \in [0,1/3)$ and $s_2 \in [0,1/3)$ that satisfy $s_1 + s_2 = .3$ the joint distribution of $\tilde{s}_1$ and $\tilde{s}_2$ is uniform over $[0,1)^2$ such that $frac(\tilde{s}_1 + \tilde{s}_2) = 0.5$ (cf. Lemma \ref{lem:cond_mask}).

\section{Conclusion}
\label{sec:dis}

In this paper, we propose a general approach (distributed and asynchronous) to ensure privacy of honest agents in any distributed average consensus protocol. The inputs of the agents are assumed to be finite real-values. The proposed approach guarantees (perfect) privacy of honest agents against passive adversarial agents if the set of adversarial agents is not a vertex cut of the underlying communication network. The only information that adversarial agents can get on the inputs of honest agents is their sum (or average). 

It is not difficult to see that the privacy protocol proposed in this paper be used for privacy in distributed computation of any function $h: \mathbb{R}^n \to \mathbb{R}$, over agents inputs $\{s_i\}$, of the following form 
\[h(s_1, \ldots, \, s_n) = g \left( \sum_{i} h_i(s_i) \right).\]
Here, $h_i: \mathbb{R} \to \mathbb{R}, \, \forall i$ and $g: \mathbb{R} \to \mathbb{R}$. We assume that the functions $h_i, \, \forall i$ are injective (one-to-one), thus privacy of $h_i(s_i)$ is equivalent to the privacy of $s_i$. Also, it is reasonable to assume that $h_i(s_i)$ is finite if $s_i$ is finite. For now, let $h_i(s_i) \in [0,1/n), \forall i$. 

Each agent first computes the effective function values $\tilde{h}_i(s_i) = frac(h_i(s_i) + a_i)$ and then uses any (non-private) distributed average consensus on these effective function values to compute $\sum_{i}\tilde h_i(s_i)$. Then, $frac\left(\sum_{i}\tilde h_i(s_i)\right) = \sum_i h_i(s_i)$ as $\sum_i a_i = 0$. Thus, each agent correctly computes the desired function value as
\[g\left( frac\left(\sum_{i}\tilde h_i(s_i)\right) \right) =  g\left( \sum_{i} h_i(s_i) \right) = h(s_1,\ldots,s_n).\]

\bibliographystyle{IEEEtran}
\bibliography{references_consensus}

\appendix

\subsection{Proof of Lemma \ref{lem:dist_a}}
\label{sub:dist}
The proof is obvious for $n = 1$. From now on, $n > 1$ and $\G$ is assumed connected.

Keep in mind that each $\{b_e\}_{e \in \E}$ is \emph{independent and uniformly distributed} in $[0,1)$. (As for any $e = \{i,j\} \in \E$, the values $r_{ij}$ and $r_{ji}$ are independent and uniform in $[0,1)$.) 


Consider a subset $\E'$ of $\E$ with $n-1$ edges such that $\mathcal{G}' = \{\V, \, \E'\}$ is connected (such $\E'$ is guaranteed to exist as $\G$ is connected). Therefore, all the $n-1$ columns of $\nabla'$ (incidence matrix of $\G'$) are linearly independent. This implies that all the points in the span (coefficients belonging to $[0,1)$) of the columns of $\nabla'$, given by \[L(\nabla') = \left( frac(\nabla' \cdot b) \, | \, b \in [0,1)^{n-1} \right),\] are equally probable as $b$ is uniformly distributed in $[0,1)^{n-1}$(note that this claim holds because all the elements of $\nabla'$ belong to $\{-1,0,1\}$). Alternately, 
\begin{align*}
    a' =  frac (\nabla' \cdot b ) = frac \left( \sum_{e \in \E'} \nabla'_{*,e} \cdot b_e \right)
\end{align*}
is uniformly distributed over $[0,1)^{n-1}$. Furthermore, combining the above with the fact that $1^T_n \cdot \nabla' = 0$ implies that $a' \in L(\nabla') \iff frac(1^T_n a' ) = 0$ (for all $a' \in [0,1)^n$). 

In case $\E' = \E$ ( or $\E$ has only $n-1$ edges), the proof concludes here. Otherwise, choose an edge $e'$ from the set of remaining edges $\E' \setminus \E$. Now, $\nabla'_{*,e'}$ can be obtained by linearly combining the columns of $\nabla'$ as following
\begin{align}
    \nabla'_{*,e'} = \sum_{e \in \E'}\mu_e \nabla'_{*,e} \label{eqn:nabla_p}
\end{align}
as $\G'$ is connected\footnote{It follows easily from the fact that there exists a path in $\G'$ between the terminal nodes of the edge $e'$ as $\G'$ is connected.}, where $\mu_e \in \{-1, 0, 1\}$ for all $e \in \E'$ . 

Define a new set of edges $\E'' = \E' \cup \{e'\}$. From \eqref{eqn:nabla_p}, each point $a''$ of $L(\nabla'')$ (span of the columns of the incidence matrix $\nabla''$ of $\G'' = \{\V, \, \E''\}$) is given as
\begin{align*}
    a'' & = frac \left( \sum_{e \in \E'}\nabla'_{*,e} \cdot b_e + \nabla'_{*,e'} b_{e'}\right)\\
    & = frac \left( \sum_{e \in \E'}\nabla'_{*,e} \cdot (b_e + \mu_e b_{e'}) \right)
\end{align*}
As the values~$\{b_e\}_{e \in \E'}$ are independent and uniform in $[0,1)$, this implies $\{ frac(b_e + \mu_e b_{e'})\}_{e \in \E'}$ is uniformly distributed over all the values in $[0,1)^{n-1}$. Hence, $a''$ is uniformly distributed over $L(\nabla')$ (and $L(\nabla'')$ is same as $L(\nabla)$).

Same as before, if $\E'' = \E$, the proof concludes. Otherwise, repeat the above procedure by considering another edge from $\E \setminus \E''$, which leads to the same conclusion. This iterative process of including edges stops when all the edges of $E$ have been considered. Ultimately, we reach the conclusion that to express $a$ is uniformly distributed in $L(\nabla')$. (Axiomatically, this also implies that $L(\nabla)$ is same as $L(\nabla')$ .) 

Hence, $a$ is uniform in $[0,1)^{n}$ subject to the constraint that $frac(1^T_n a) = 0$ (or $frac(\sum_i a_i) = 0$), as any point $a \in L(\nabla') \iff frac(1^T_n a ) = 0$.

\subsection{Proof of Lemma \ref{lem:cond_mask}}
\label{sub:cond}
Let $\tilde{S}$, $S$ and $A$ represent the random vectors of the agents' effective inputs, true inputs and masks, respectively. $\G$ is assumed connected. For a random variable (or vector) $S$, $f_{S}(s)$ denotes its probability density at $s$.

As $\tilde{s}_i = frac (s_i + a_i)$ and $s_i, a_i$ are independent, we have
\begin{align*}
    &f_{\tilde{S}}\left(\tilde{s} | S =  s \right) = f_{A}\left(A = frac (\tilde{s} - s) \right)
\end{align*}
If $frac (\sum_i \tilde{s}_i) = \sum_i s_i $ then $frac(\tilde{s} - s)$ belongs to $L(\nabla)$. Thus, from Lemma \ref{lem:dist_a},
\[f_{\tilde{S}} \left(\tilde{s} | S = s\right) = 1\] 
for all the values $\tilde{s}$ in $[0,1)^{n}$ that satisfy $frac (1^T_n \tilde{s}) = 0$, when $\G$ is connected.

\subsection{Proof of Theorem \ref{thm:perf}}
\label{sub:perf}
Let $\G_{\H} = \{\H, \, \E_{\H}\}$ be the graph of honest agents (and edges incident to only honest agents) and $\nabla_{\H}$ be its \emph{incidence matrix}. For a random variable (or vector) $S$, $f_{S}(s)$ denotes its probability density at $s$.

The \emph{view} of adversarial agents in $\C$ consists of honest agents' effective inputs $\tilde{s}_{\H}$ after the first phase of our protocol (considering the ``worst-case" scenario where agents' can acquire all the inputs through their internal states in $\Pi$) and the values $\{b_e\}_{e \in \E_{\C}}$. Therefore, 
\[\view_{\C}(s) = \{\tilde{s}_{\H}, \,\{b_e\}_{e \in \E_{\C}}\}\]
given the inputs $s \in [0,1/n)^n$.

Thus, we prove that the joint probability distribution of $\tilde{s}_{\H}$ and $\{b_e\}_{e \in \E_{\C}}$ is the same for any two sets of true inputs $s, s'$, that satisfy $s_{\C} = s'_{\C}$ and $\sum_{i \in \V}s_i = \sum_{i \in \V}s'_i$, when $\G_{\H}$ is connected. 

We have,
\begin{align*}
    a_i = frac\left( frac\left( \sum_{e \in \E_{\H}}\nabla_{i,e} b_e \right) + frac \left(\sum_{e \in \E_{\C}}\nabla_{i,e} b_e \right) \right)
\end{align*}

The values $\{ frac(\sum_{e \in \E_{\H}}\nabla_{i,e} b_e)\}_{i \in \H}$ lie in the span of $\nabla_{\H}$, denoted by $L(\nabla_{\H})$. Therefore, $\{ frac( \sum_{e \in \E_{\H}}\nabla_{i,e} b_e) \}_{i \in \H}$ is uniformly distributed over $[0,1)^{|\H|}$ subject to $frac(\sum_{i \in \H}(\sum_{e \in \E_{\H}}\nabla_{i,e} b_e)) = 0$ when $\G_{\H}$ is connected (cf. Lemma \ref{lem:dist_a}). \\
Thus, it is clear that the masks $\{a_i\}_{i \in \H}$ are uniformly distributed in $[0,1)^{|\H|}$ subject to \\ $frac(\sum_{i \in \H} a_i) = frac(-\sum_{i \in \C} a_i)$ when $\G_{\H}$ is connected (given the values of $\{b_{e}\}_{ e \in \E_{\C}}$).\\
Note that random variables $\{b_e\}_{e \in \E_{\H}}$ are uniformly and independently distributed in $[0,1)$, given the values of $\{b_e\}_{e \in \E_{\C}}$, and $a_{i} = frac(\sum_{e \in \E_{\C}}\nabla_{i,e}b_e)$ for every $i \in \C$. Thus using Lemma \ref{lem:cond_mask} above implies, ($\tilde{S}_{\H}$ denotes the random vector of honest agents' effective inputs $\tilde{s}_{\H}$) 
\begin{align}
    f_{\tilde{S}_{\H}}\left(\tilde{s}_{\H} | s_{\H}, \, \{b_{e}\}_{ e \in \E_{\C}} \right) = 1 \label{eqn:f_cond}
\end{align}
for all the values $\tilde{s}_{\H}$ in $[0,1)^{|\H|}$ that satisfy 
\begin{align*}
    frac \sum_{i \in \H} \tilde s_i  = frac\left(\sum_{i \in \V} s_i - \sum_{i \in \C} \tilde s_i \right)  
\end{align*}
when $\G_{\H}$ is connected. 

Combining \eqref{eqn:f_cond} with the fact that the random variables $\{b_e\}_{e \in \E}$ are independent to the true inputs $s$ implies 
\begin{align*}
    f_{\view_{\C}(s)}\left(\{\tilde{s}_{\H}, \{b_e\}_{e \in \E_{\C}}\}\right) \equiv f_{\view_{\C}(s')}\left( \{\tilde{s}_{\H}, \{b_e\}_{e \in \E_{\C}}\}\right)
\end{align*}
for all $s, s' \in [0,1/n)^n$ such that $s_{\C} = s'_{\C}$ and $\sum_{i \in \V}s_i = \sum_{i \in \V}s'_i$ when $\G_{\H}$ is connected. 





\end{document}